\documentclass[times]{aa}
\usepackage{psfig}
\usepackage{astron}

\begin{document}

\def\aap{A\&A}
\def\apj{ApJ}
\def\apjl{ApJL}
\def\mnras{MNRAS}

\def\crp{CRp }
\def\cre{CRe }
\def\eps{\varepsilon}

\def\e{{\rm e}}
\def\p{{\rm p}}

\thesaurus{ 
03(02.13.1;        
02.13.2;        
02.18.5;        
11.09.3;        
11.03.1;        
13.18.2)        
}

\title{Radio Halos of Galaxy Clusters from Hadronic Secondary Electron
Injection in Realistic Magnetic Field Configurations}
\titlerunning{Radio Halos from Hadronic Secondary Electrons}
\author{Klaus Dolag \and Torsten A. En{\ss}lin}\authorrunning{Klaus
Dolag \and Torsten A. En{\ss}lin}
\institute{Max-Planck-Institut f\"{u}r Astrophysik,
Karl-Schwarzschild-Str.1, 85740 Garching, Germany}
\maketitle

\begin{abstract}
We investigate the possibility that radio halos of clusters of
galaxies are caused by synchrotron emission of cosmic ray electrons
(CRe), which were produced by cosmic ray protons (CRp) interacting hadronically 
with the intra-cluster medium (ICM) protons. We perform cosmological magneto-hydrodynamics
(MHD) simulations to obtain a sample of ten magnetized galaxy
clusters. They provide
realistic models of the gas and magnetic field distribution, needed to
predict the \cre production rates, their cooling, and their
synchrotron emissivity.
We assume a \crp population within the ICM with an energy density 
which has a constant ratio to thermal energy density. 
This ratio is adjusted in such a way that one of the simulated 
clusters reproduces the radio luminosity of the radio halo of the Coma
cluster of galaxies. Our model exhibits the observed low degree of
radio polarization and has a similar radial emission profile as the
Coma cluster. We provide
estimates for the expected gamma ray and neutrino flux. The
necessary CRp/thermal energy ratio is 4 ... 14~\%~${(E_{\rm
p,min}/{\rm GeV})}^{-0.375}$ (for the range of magnetic field strengths
suggested by Faraday measurements), where $E_{\rm p,min}$ is the lower
kinetic energy cutoff of the \crp with spectral index $\alpha_{\rm p}
\approx 2.375$. Assuming this ratio to be the same in the whole set of
simulated clusters a $T_\mathrm{x}-L_\nu$ relation is predicted which
follows the observed relation well.
\end{abstract}


\section{Introduction\label{sec:intro}}

Some clusters of galaxies contain large, Mpc-sized regions of diffuse
radio emission, the so called cluster radio halos and cluster radio
relics. Although known for decades, these sources receive increasing attention
due to the improved detection sensitivities and their important
theoretical implications for the non-thermal content of the ICM.  For
reviews see Valtaoja
\cite*{1984A&A...135..141V}, Jaffe
\cite*{1992cscg.conf..109J}, Feretti \& Giovannini
\cite*{1996IAUS..175..333F}, 
Feretti
\cite*{Feretti.Pune99}, En{\ss}lin \cite*{Pune99}, and Giovannini et
al. \cite*{1999NewA....4..141G}.

Whereas the cluster relic sources are nowadays widely believed to
trace the presence of large shock waves in the ICM
\cite{1998AA...332..395E,1999ApJ...518..603R,ensslin2000b} the origin
of radio halos is less clear.
The lifetime of the radio emitting \cre is much shorter than any
reasonable transport time over cluster scales. This means that the
\cre need to be continuously re-energized, or freshly injected on a
cluster-wide scale.
The latter would happen if the \cre result from the decay of charged
pions produced in hadronic interaction of energetic \crp distributed
throughout the cluster volume
\cite{1980ApJ...239L..93D,1982AJ.....87.1266V,1984A&A...135..141V,1987A&A...182...21S,1999APh....12..169B}.
The \crp have lifetimes in the IGM of the order of a Hubble time
\cite{1997ApJ...487..529B,1997ApJ...477..560E}, giving them enough
time to diffuse away from their source and to maintain a cluster wide
distribution. 

It is the purpose of this work to demonstrate that a simple model for
hadronic electron injection in a realistic magnetic field
configuration leads to radio halos which reproduce several
observations: the profile of the radialy decreasing radio emission, the low radio
polarization, the radio to X-ray surface brightness correlation, and
the cluster radio halo luminosity-temperature relation. We do not
attempt to reproduce all observational information on cluster radio
halos. For example the observed spatial spectral index variations of
individual clusters
\cite{1993ApJ...406..399G,1997A&A...321...55D}
are not reproduced by the presented model. It is our goal to investigate
the observational consequences of a conceptually simple
model. Sophisticated modifications, which fine-tune the model to the
data, are principally possible.  But first an
understanding of idealized models is most helpful. The
assumptions of our model are:
\begin{enumerate}
\item \cre are secondary particles of hadronic interactions of \crp
with the background protons of the ICM. The origin of the \crp is not
specified here, but a likely explanation is particle acceleration in
shock waves of cluster merger events and in accretion shocks of the
clusters \cite{1998APh.....9..227C}. ICM \crp might also migrate from
radio plasma of radio galaxies into the ICM
\cite{1984A&A...135..141V,1997ApJ...477..560E,Ringberg99,1999APh....12..169B}
or result from supernova driven galactic winds
\cite{1996SSRv...75..279V}.
\item The energy density in \crp is assumed to be proportional to the
energy density of the thermal ICM gas. This is reasonable if the
thermal gas and the \crp were energized in the same shock waves of
cluster mergers, since a constant fraction of energy dissipated in the shock
should go into the \crp population.  Previous work ignored the spatial profiles
\cite{1980ApJ...239L..93D,1982AJ.....87.1266V} or assumed one single
central point source from which the \crp diffused away, resulting in a
centrally very peaked \crp profile
\cite{1999APh....12..169B}.
\item The \crp spectral index is independent of position or cluster
and therefore the radio spectral index is constant in this
model. Although this is certainly an oversimplification since e.g. the radio halo of
the Coma cluster does show spectral index variations, it allows to
predict radio halo statistics.
\item No significant re-acceleration or particle diffusion of the \cre
is occurring.
\item The radio spectrum is assumed to be quasi-stationary, since the
electron cooling time is short and a stationary \cre population is
rapidly established. The spectrum of this population results from the
shape of the proton spectrum, which is assumed to be a single
power-law, and the synchrotron and inverse Compton cooling. Hence
electron and radio spectra are also power laws.
\item The magnetic field configuration results from 
amplification of pre-cluster seed fields as simulated by Dolag et
al.~\cite*{1999A&A...348..351D}. We therefore investigate very
inhomogeneously magnetised clusters. Although the importance of
realistic field configurations for the appearence of radio halos was
stressed by Tribble
\cite*{1991MNRAS.253..147T}, even recent work on halos of hadronic
origin assumed spatially constant field strength.

\end{enumerate}

\section{MHD Cluster Formation Simulation}

\begin{figure}[t]
\begin{center}
\psfig{figure=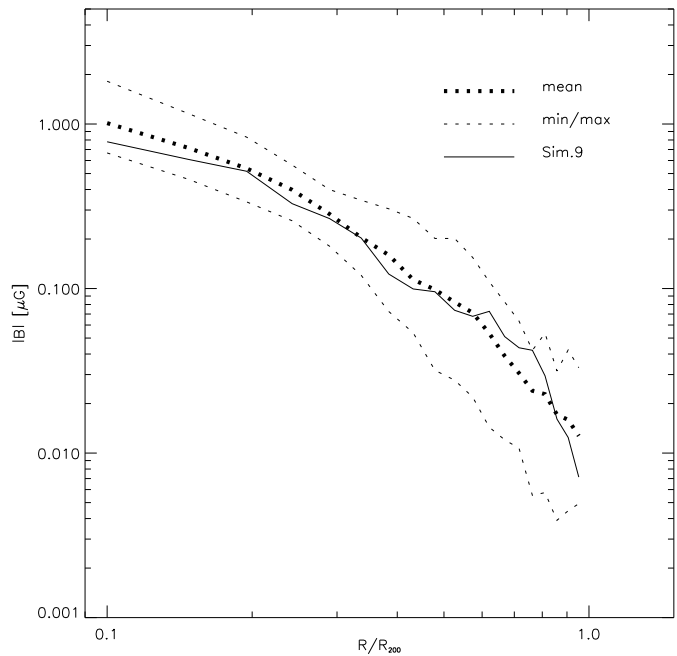,width=0.48 \textwidth,angle=0}
\end{center}
\caption[]{\label{fig:vgl.dprof} 
  The radial profile of the magnetic field strength of the simulated
  clusters is shown in units of the virial radius $R_{200}$.
  The solid line represents the field strength in the cluster
  Sim.9. The thick dotted line is the profile averaged over all ten
  simulations, the thin dotted lines are the minimum and maximum
  values found in the different simulations.}

\end{figure}

\begin{figure}[t]
\begin{center}
\psfig{figure=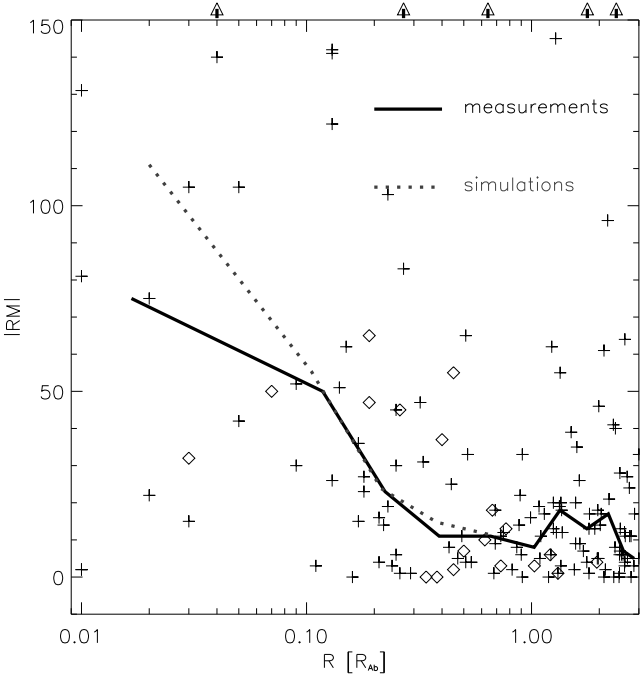,width=0.48 \textwidth,angle=0}
\end{center}
\caption[]{\label{fig:vgl.dbl} 
  Comparison of synthetic Faraday-rotation measurements with a sample
  of measurements in Abell clusters. The absolute values of Faraday
  rotation measurements \cite{1991ApJ...379...80K} vs.~radius in units
  of the Abell radius are shown. Obviously, the dispersion increases
  towards the cluster center. The solid curves mark the median of the
  measurements, the dotted curve is the median obtained from the
  simulated cluster sample. Some measurements have Faraday
  measurements which are outside the region shown (marked by the
  arrows on top of the plotting region). The median was chosen because
  it is not influenced by such values. The flattening of the
  observations towards the cluster center seems to be an artefact in
  the data. A comparison with newer data \cite{Clarke,clarke2000sub}
  shows that the inner-most bin is also perfectly in agreement with
  the simulations. }
\end{figure}

We used the cosmological MHD code described in Dolag et 
al.~\cite*{1999A&A...348..351D} to
simulate the formation of magnetised galaxy clusters from an initial
density perturbation field. The evolution of the magnetic field is 
followed from an initial seed field. This field is amplified by the 
compression during the cluster collapse. Merger events and sheer flows, which are
very common in the cosmological environment of large-scale structure, 
lead to Kelvin-Helmholtz instabilities. They further increase the
field strength by a large factor. In order to reproduce the present
$\mu$G fields an initial nG field strength is required.

\subsection{GrapeSPH}
The code combines the merely gravitational interaction of a
dark-matter component with the hydrodynamics of a gaseous
component. The gravitational interaction of the particles is evaluated
on GRAPE boards, while the gas dynamics is computed in the SPH
approximation. It was also supplemented with the magneto-hydrodynamic
equations to trace the evolution of the magnetic fields which are
frozen into the motion of the gas because of its assumed ideal
electric conductivity. The back-reaction of the magnetic field on the
gas is included.  Extensive tests of the code were successfully
performed and described in the previous paper. $\vec\nabla \cdot
\vec{B}$ is always negligible compared to the magnetic field divided
by a typical length scale. The code also assumes the ICM to be an
ideal gas with an adiabatic index of $\gamma=5/3$ and neglects
cooling. The surroundings of the clusters are dynamically important
because of tidal influences and the details of the merger history. In
order to account for this the cluster simulation volumes are
surrounded by a layer of boundary SPH particles in order to
represent accurately the sources of the tidal fields in the cluster
neighborhood.  For details of the code, the models and the obtained
magnetic field structure see Dolag et al. (1999; in preparation).

\subsection{Initial conditions}
For an accurate calculation of the synchrotron emission the resolution
of the magnetic field is crucial. We therefore perform simulations in
a CDM Einstein-de-Sitter cosmology ($\Omega_{\mathrm m}^0=1.0$,
$\Omega_\Lambda^0=0$, $H_{0} = 50\,{\rm km\,s^{-1}\,Mpc^{-1}}$), with
twice the mass resolution of the simulation described in Dolag
(1999). We simulate a set of ten different realisations which result
in clusters of different final masses and different dynamical states
at redshift $z=0$.

As shown in Dolag 1999, the initial field structure is
irrelevant because the final field structure is mostly determined by
the dynamics of the cluster collapse. The initial field strengths in
this simulations where chosen so that the final field strength
reproduces the observed Faraday rotation data collected by Kim et
al.~\cite*{1991ApJ...379...80K}.  The central values of the ten
clusters vary between $0.5\mu$G and $2\mu$G, whereas Sim.9 reaches
values of $\approx 1\mu$G.  The simulated magnetic field strength
drops by two orders of magnitudes from the central part out to the
virial radius (see Fig. \ref{fig:vgl.dprof}).

Given the large uncertainties a field strength twice as high may
also be consistent with the Faraday data.  Therefore we also give
results for the extrapolated case of field strength twice as high.
We note
that neither the simulation, nor the Faraday rotation measurements can
exclude the existence of stronger small-scale ($\le 1$~kpc) magnetic
field components. Here, we assume only the presence of large-scale
($\ge 20$~kpc) and relatively weak fields. A comparison of the
synthetic Faraday rotation measurements with the observations is shown
in Fig. \ref{fig:vgl.dbl}.

\section{Hadronic Secondary Electron Injection\label{hadronic}}

A \crp population may be well described by a power-law in kinetic
energy $E_{\rm p}$
\begin{equation}
f_\p (\vec{x} ,E_\p) \,dE_\p\,dV = \frac{N_\p(\vec{x})}{\rm GeV}\, \left(
\frac{E_\p}{\rm GeV} \right)^{-\alpha_\p}\,dE_\p\,dV\,.
\end{equation}
We assume a lower cutoff at $E_{\rm p, min} \approx 1 \,{\rm
GeV}$. The contribution of He (assuming a 24\% mass fraction) to the
cosmic rays and thermal gas is implicitly taken into account in our
calculations by replacing proton densities by the correct nucleon
densities without further notification.  We choose the normalization
$N_\p(\vec{x})$ so that the (kinetic) \crp energy density $\eps_{\rm
\crp} (\vec{x})$ is a constant fraction of the thermal energy density
$\eps_{\rm th} (\vec{x})$ of the ICM.
\begin{equation}
\eps_{\rm \crp} =  X_{\rm \crp}\, \eps_{\rm th} = \frac{N_{\rm
p}(\vec{x})}{\alpha_\p -2} \, \left(
\frac{E_{\rm p,min}}{\rm GeV} \right)^{2-\alpha_\p}\,{\rm GeV}
\end{equation}
Since the spectral index is sufficiently steep in our model
($\alpha_\p \approx 2.4$) the upper cutoff was set to infinity.

The \crp interact hadronically with the background gas and produce 
pions. The charged pions decay into secondary electrons (and neutrinos)
and the neutral pions into gamma-rays:
\begin{eqnarray}
  \label{eq:pp}\nonumber
  \pi^\pm &\rightarrow& \mu^\pm + \nu_{\mu}/\bar{\nu}_{\mu} \rightarrow
  e^\pm + \nu_{e}/\bar{\nu}_{e} + \nu_{\mu} + \bar{\nu}_{\mu}\nonumber\\
  \pi^0 &\rightarrow& 2 \gamma\nonumber
\end{eqnarray}
The resulting electron injection spectrum $q_\e(\vec{x}, E_\e) \,
dE_\e\,dV$ peaks at the energy of $m_{\pi}\,c^2/4 \approx 35\,{\rm
MeV}$ due to rough energy equipartition between electrons and
neutrinos in the charged pion decay \cite{1994A&A...286..983M}. The
radio emission is produced by electrons with an energy of several GeV
(depending on observing frequency and magnetic field strength), which
have to result from protons of even higher energies. For this energy
range the electron production spectrum can be calculated following
Mannheim \& Schlickeiser
\cite*{1994A&A...286..983M} to be
\begin{equation}
q_\e(\vec{x},E_\e) \approx 2^6\,\sigma_{\rm pp}\,c\,n_{\rm
p}(\vec{x})\, \frac{N_\p(\vec{x})}{\rm GeV} \, \left(
\frac{24\,E_\e}{\rm GeV}
\right)^{-\frac{4}{3}(\alpha_\p -\frac{1}{2})}\!\!\!\!\!\!,
\end{equation}
where $\sigma_{pp} = 32\,{\rm mbarn}$ is the inelastic p-p cross
section, and $n_{\rm p}$ is the target proton density . We
note, that the neutrino production spectrum is $q_\nu(\vec{x}, E_\nu)
= 3\,q_\e(\vec{x},E_\nu)$, and the $\pi^0$-decay induced gamma-ray
spectrum is $q_\gamma(\vec{x}, E_\gamma) =
3\,q_\e(\vec{x},E_\gamma/4)/(16 \,\alpha_\p -8)$ for $E_\gamma \gg
m_{\pi^0}\, c^2/2 \approx 68\, {\rm MeV}$.

The steady-state \cre spectrum is shaped by the injection and
cooling process and is governed by
\begin{equation}
\frac{\partial }{\partial E_\e} \left( \dot{E_\e}(\vec{x},E_\e) f_\e
(\vec{x},E_\e) \right) = q_\e(\vec{x}, E_\e)\,.
\end{equation}
For $\dot{E_\e}(\vec{x},p) < 0$ this equation is solved by
\begin{equation}
f_\e (\vec{x},E_\e) = \frac{1}{|\dot{E_\e}(\vec{x},E_\e)|} \int_{E_\e}^\infty
\! dE_\e'  q_\e(\vec{x}, E_\e')\,.
\end{equation}
The cooling of the radio emitting \cre is dominated by synchrotron and
inverse Compton losses giving
\begin{equation}
- \dot{E_\e}(\vec{x},E_\e) = \frac{4\,\sigma_T\, c}{3\,m_\e^2\,c^4} \left(
\frac{B^2 (\vec{x})}{8\,\pi} + \frac{B^2_{\rm CMB}}{8\,\pi}
\right)\,E_\e^2\,,   
\end{equation}
where $B(\vec{x})$ is the local magnetic field strength and $B^2_{\rm
CMB}/(8\,\pi)$ is the energy density of the cosmic microwave
background expressed by an equivalent field strength. 
The  \cre have therefore a power-law spectrum
\begin{equation}
f_\e (\vec{x},E_\e) = \frac{N_\e(\vec{x})}{\rm GeV}
\,\left( \frac{E_\e}{\rm GeV} \right)^{-\alpha_\e}\,,
\end{equation}
with $\alpha_\e = \frac{4}{3}\,\alpha_\p + \frac{1}{3}$ and
\begin{equation}
N_\e(\vec{x}) = \frac{2^7\,3^2 \,\pi\,\, 24^{-\frac{4}{3}(\alpha_\p -
\frac{1}{2})}}{4\,\alpha_\p - 5}   \,
\frac{\sigma_{\rm pp}\, m_\e^2\,c^4}{\sigma_T \,{\rm GeV}} \,
\frac{n_\p(\vec{x})\, N_\p(\vec{x})}{B^2(\vec{x}) + B^2_{\rm
CMB}}\,.
\end{equation}
The radio emissivity $j_\nu$ at frequency $\nu$ and per steradian is
given by
\begin{equation}
   j_\nu=c_2(\alpha_\e)\, N_\e(\vec{x}) \, 
   B_\perp(x)^{\frac{\alpha_\e+1}{2}}
   \left(\frac{\nu}{c_1}\right)^{-\frac{\alpha_\e -1}{2}}
\end{equation}
\cite{1999asfo.book.....L} with $c_1=3\,e\,{\rm GeV^2}/(2\,\pi\, m_\e^3\,c^5)\,$,
\begin{equation}
   c_2(\alpha_\e)=\frac{\sqrt{3}}{16\pi}\frac{e^3}{m_\mathrm{e}c^2}
               \frac{\alpha_\e+\frac{7}{3}}{\alpha_\e+1}
               \Gamma\left( \frac{3\alpha_\e-1}{12}\right)
               \Gamma\left( \frac{3\alpha_\e+7}{12}\right),
\end{equation}
and $B_\perp(\vec{x})$ being the magnetic field component within the image
plane. Integration of $j_\nu (\vec{x})$ along the $z$-direction gives
the surface brightness of the radio halo.

Thus we find that the radio emissivity as a function of position in
the hadronic secondary electron model scales with the ICM properties
as
\begin{equation}\label{eq:jnu}
j_\nu \sim  X_{\rm \crp}\,n_\p\,\eps_{\rm th}
\frac{B_\perp^{\alpha_{\nu}+1}}{B^{2} + B^2_{\rm CMB}}\,
\nu^{-\alpha_{\nu}}\,,
\end{equation}
where $\alpha_{\nu} = (\alpha_\e-1)/2 = (2\,\alpha_{\rm
p}-1)/3$. It follows that the emissivity of strong field regions ($B^2
\gg B^2_{\rm CMB}$) is nearly independent of the magnetic field
strength, whereas in weak field regions ($B^2 \ll B^2_{\rm CMB}$) it
is very sensitive to it. The factor $n_\p\,\eps_{\rm th}$ is nearly
identical to the X-ray emissivity $l_X \sim n_{\rm
p}^2\,(kT_\e)^{1/2}$ so that for a radially decreasing magnetic field
strength one always gets a radio luminosity which declines more
rapidly than the X-ray emission:
\begin{equation}\label{eq:jnulx}
\frac{j_\nu}{l_X}  \sim  X_{\rm \crp}\,(kT_{\rm e})^{1/2}
\frac{B_\perp^{\alpha_{\nu}+1}}{B^{2} + B^2_{\rm CMB}}\,
\nu^{-\alpha_{\nu}}\,,
\end{equation}
The normalization constant of Eq. \ref{eq:jnu} depends on the assumed
scaling between \crp and thermal energy density. We fix this relation
by comparing Coma to a similar simulated cluster (Sim.~9) and assume it
to hold for all clusters in order to predict their radio luminosity.

\begin{figure}[t]
\begin{center}
\psfig{figure=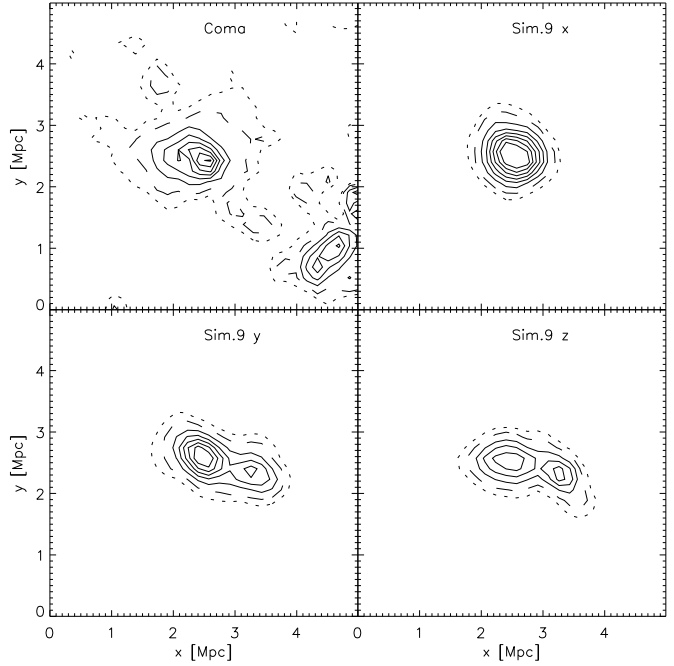,width=0.48\textwidth,angle=0}
\end{center}
\caption[]{\label{fig:big9.map} The upper left panel shows the
Effelsberg radio map of the radio halo in the Coma cluster from Deiss
et al. \cite*{1997A&A...321...55D}. The central radio halo, the
peripheral cluster radio relic 1253+275, and the so called radio
bridge between them are visible.  The other panels show synthetic
radio maps of a simulated cluster (Sim.~9) at 1.4GHz in three
spatial projections. For brtter comparison they are smoothed to the 
resolution of the radio observation. Contour levels are
0.1, 0.2, 0.4, 0.6, 0.8, 1.0, 1.2, 1.4 $\mathrm{mJy}/\mathrm{arcmin}^2$.}
\end{figure}

\begin{figure}[t]
\begin{center}
\psfig{figure=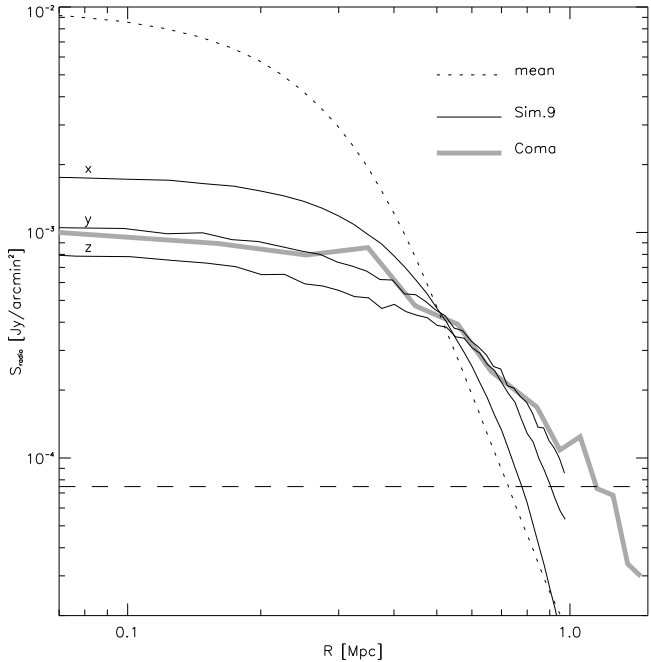,width=0.48\textwidth,angle=0}
\end{center}
\caption[]{\label{fig:big9.prf} 
  The radial profile (centered on the emission weighted maximum) of the Coma cluster (thick line) is compared to
  the profiles obtained from the three projections of the simulation 
  shown in Fig.~\ref{fig:big9.map} (thin lines). The dotted line represents
  the mean profile over all ten simulated clusters. The horizontal
  line marks a crude estimate of the noise enclosed in
  the observations.}
\end{figure}

\section{Synthetic Radio Maps}

To obtain the synthetic radio maps we calculate all quantities needed
in a $5\,\mathrm{Mpc}^3$ box containing $200^3$ grid cells, centered on
the cluster position and distribute the CRe as described before.
From this data cube we calculate the maps by projecting the emission along the 
three spatial directions. As the radio emission depends on the
angle between the magnetic field and the observing direction, this
leads to different total radio luminosities in different observing
directions. As the magnetic field in the simulations is nearly
isotropic, the differences are small. Additionally projection effects
of substructure in the underlying density distribution can lead to
different radial profiles as seen in Fig. \ref{fig:big9.prf}.

As a reference point we chose one simulated cluster (Sim.~9), which
has an emission weighted temperature (9.4 keV) and a velocity
dispersion (1200 km/s derived from the virial mass and radius)
comparable to the values observed in the Coma cluster of galaxies (9
keV (Donelly et al., 1999\nocite{1999ApJ...513..690Dshort1}), 1140 km/s
\cite{1993AJ....106.1301Z}).  This cluster is in a post-merger
stadium, like Coma \cite{1994ApJ...427L..87B,2000AJ....119..580B}.

We assume $\alpha_p = 2.375$, which gives $\alpha_\nu = 1.25$, typical
for cluster radio halos, and adjust $X_{\rm CRp}$ to $0.14 \,{(E_{\rm
p,min}/{\rm GeV})}^{-0.375}$ so that this cluster reproduces the
observed total 1.4 GHz radio emission of Coma. In the case of a
twice the field strength $X_{\rm CRp}$ needs to be $ 0.04 \,{(E_{\rm
p,min}/{\rm GeV})}^{-0.375}$. Note that in this case the magnetic
fields and the CRp are in equipartition for $E_{\rm p,min} = 1\, {\rm
GeV}$.

The synthetic radio maps are shown in Fig.~\ref{fig:big9.map} in
comparison with the radio map of Coma.   In the calculation of the radial radio
emission profile of Coma the so called {\it radio bridge} of Coma (see
Fig. \ref{fig:big9.map}, top-left panel at $x=3.5$ Mpc, $y=1.5$ Mpc)
is included.  This can be justified because a similar bridge is also
visible in the X-ray emission, which suggests that the radio bridge
results from ICM substructure. This substructure is believed to be the
core of a smaller merging cluster which already passed the core
region of Coma \cite{1994ApJ...427L..87B,2000AJ....119..580B}.  The
radio relic in Coma (see Fig. \ref{fig:big9.map}, top-left panel at
$x=4.5$ Mpc, $y=1$ Mpc) does not affect the comparison of the radio
profiles due to its peripheral location.

Although we cannot expect this simulated cluster to match the
morphology of the radio halo of Coma exactly, the radial emission
profiles in the $y$- and $z$-projection are quite comparable
(Fig.~\ref{fig:big9.prf}).  In the $x$-projection the two sub-lumps
are projected on top of each other. This leads to a steeper radial
radio profile (Fig. \ref{fig:big9.prf}).  The second simulated
cluster with flat radio emission profile is also in a post-merger
stadium with substantial substructure.

The large extension of low level radio emission of Coma is not
reproduced by the model. As the simulation has its highest resolution
in the center, the simulation might underestimate the small scale
magnetic fields, and therefore the radio emission, in the outer
regions. This could explain the less extended synthetic radio
halo. Another possibility is that e.g. diffusion established a more
extended CRp profile than it is assumed in this work.

The observed radio halos do not show any radio polarization. We
estimate for our simulated field configuration the expected degree of
radio polarization following the treatment of internal Faraday
rotation of Burn \cite*{1966MNRAS.133...67B}. 
The synthetic polarization maps has peak values of the order of
$10^{-6}$. This undetectable signal would be further reduced by beam
depolarization due to the small-scale polarization pattern.

\begin{figure}[t]
\begin{center}
\psfig{figure=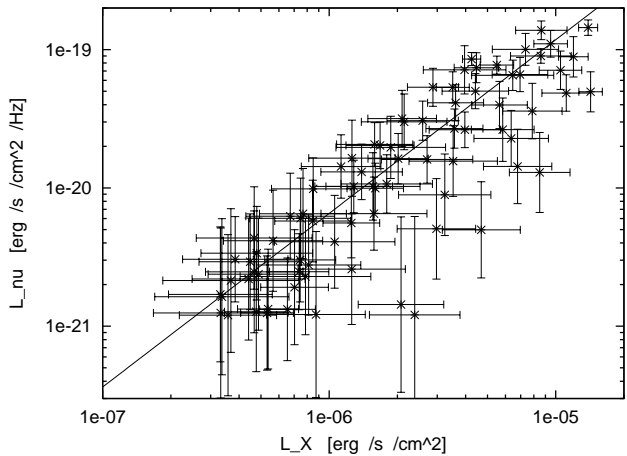,width=0.5\textwidth,angle=0}
\end{center}
\caption[]{\label{fig:tst2} Point by point radio to X-ray comparison of Sim.~9.}
\end{figure}

It is difficult to compare morphological different clusters of
galaxies. But Eq. \ref{eq:jnulx} indicates a close relation between
radio and X-ray emissivities within the hadronic halo model. This
should be approximatively conserved in the projected fluxes, which
allows for a less morphology-dependent comparison of different real or
simulated clusters. For Sim.~9 we compare X-ray and radio surface
brightness point by point. In order to process the data in the same
way as real observational data we average over 250 kpc boxes and use
the RMS as an error estimate (plus systematic errors of $\delta L_X =
10^{-7}\,{\rm erg\, s^{-1}\, cm^{-2}}$ and $\delta L_\nu =
10^{-21}\,{\rm erg\, s^{-1}\, cm^{-2}\,Hz^{-1}}$).  The results are
given in Fig. \ref{fig:tst2}. For instrumentally detectable regions
($L_\nu > \delta L_\nu$, $L_X > \delta L_X$) the radio and X-ray
emissivity can be related via a power law $L_\nu = a \, L_X^b $, where
$a = 2.23\cdot 10^{-13}$ and $b = 1.26$ in cgs units.  Govoni et
al. \cite*{govoni2000sub} give for four clusters a point by point
radio to X-ray count rate correlation. Their values ($b = 0.64 \pm
0.07$ for Coma, $b= 0.82\pm 0.04$ for A2319, $b= 0.94\pm 0.04$ for
A2255, $b = 0.99\pm 0.05$ for A2744) are significantly smaller, which
is also supported by the visual impression of the more concentrated
simulated halo in comparison to Coma (Fig. \ref{fig:big9.map}).  This
indicates that the radial \crp profile should be more extended than
the thermal energy of the ICM, as we have assumed in this work, if
radio halos are of hadronic origin.

\section{Gamma-Rays and Neutrinos}

The luminosity of the simulated cluster Sim.~9 in neutrinos and
gamma-rays is $Q_{\nu_{\e,\mu}}(E_\nu) = 5.5 \cdot 10^{45} \,
E_\nu^{-2.5}/{\rm GeV^{-1.5}}$ and $Q_{\gamma}(E_\gamma) = 5.9 \cdot
10^{45} \,E_\gamma^{-2.5}/{\rm GeV^{-1.5}}$. This leads to a gamma ray
flux of $\le 5.2 \cdot 10^{-8} {\, \rm cm^{-2} \, s^{-1}}$ above 100
MeV ($\le$ since the spectrum is significant below the above power law
near $E_\gamma = 100$ MeV). This is comparable to the upper limit $ 4
\cdot 10^{-8} {\, \rm cm^{-2} \, s^{-1}} $ for the flux from Coma
measured by EGRET (Sreekumar et al., 1996\nocite{1996ApJ...464..628Sshort}). In the
case of twice the field strength one finds $Q_{\nu_{\e,\mu}}(E_\nu) = 1.5
\cdot 10^{45} \, E_\nu^{-2.5}/{\rm GeV^{-1.5}}$, and
$Q_{\gamma}(E_\gamma) = 1.6 \cdot 10^{45} \,E_\gamma^{-2.5}/{\rm
GeV^{-1.5}}$, which leads to a flux above 100 MeV of $\le 1.4 \cdot
10^{-8} {\, \rm cm^{-2} \, s^{-1}}$ well below the EGRET limit. For
comparison with gamma ray and neutrino flux predictions for clusters
of galaxies see V\"olk et al. \cite*{1996SSRv...75..279V}, En{\ss}lin
et al. \cite*{1997ApJ...477..560E}, Colafrancesco \& Blasi
\cite*{1998APh.....9..227C}, and Blasi \cite*{1999ApJ...525..603B}.

\section{The Temperature-Radio-Luminosity Relation}

\begin{figure}[t]
\begin{center}
\psfig{figure=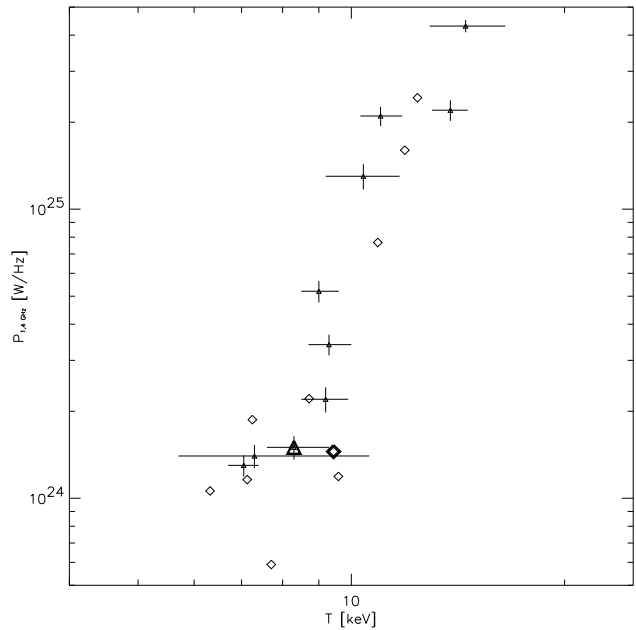,width=0.48\textwidth,angle=0}
\end{center}
\caption[]{\label{fig:lxtnu} 
  Temperature-radio luminosity relation. The data points with the error
  bars are taken from Liang \cite*{Ringberg99h}, the diamonds are the
  simulated clusters. Bigger symbols mark Coma and Sim.~9.}
\end{figure}

Clusters with radio halos exhibit a strong correlation between the
radio luminosity (at 1.4 GHz: $L_{\rm 1.4 GHz}$) and the ICM
temperature ($T_X$) \cite{Ringberg99h,Ringberg99Colafrancesco}. This
is expected because of the known correlations of cluster size and
energy content with temperature
\cite{1997ApJ...491...38M,1999A&A...349..435S,1999ApJ...517..627M}. But
the shape of the $T_X$--$L_{\rm 1.4 GHz}$ relation is an important
touchstone of every radio halo theory. It can be seen in
Fig.~\ref{fig:lxtnu} that the hadronic halo model reproduces the
observed relation if a cluster independent $\eps_{\rm CRp}/\eps_{\rm
th}$ ratio is assumed.

\section{Conclusion}
A set of simulated clusters of galaxies in a CDM EdS cosmology,
including fully MHD simulated magnetic field structures, was used to
test a conceptual simple hadronic secondary electron model for
radio halo formation of clusters. The seed magnetic fields were chosen so that
the final magnetic fields reproduce the observed Faraday rotation
measurements.  It was assumed further, that in all clusters a spatially
constant ratio between the cosmic ray proton and the thermal energy
density exists. We normalized this ratio by comparison of a simulated
radio halo of a Coma like cluster in our sample to the observed radio
halo of the Coma cluster. With the further simplifying assumption of
an unique power law proton energy spectrum, the resulting radio halo,
gamma ray and neutrino emission could be calculated.  The main results
are:
\begin{enumerate}
\item A cosmic ray proton to thermal gas energy ratio of 4 ... 14~\%~${(E_{\rm
p,min}/{\rm GeV})}^{-0.375}$ seems to be required to reproduce the
total flux at 1.4 GHz of the Coma cluster with the simulated Coma like
cluster. $E_{\rm p,min}$ is the lower kinetic energy cutoff of the
protons with assumed spectral index $\alpha_{\rm p}
\approx 2.375$. 
\item The radial profile of the simulated Coma like cluster, is
comparable to that of Coma.
\item But a detailed point by point radio to X-ray comparison, reveals
a relation ($L_{\rm 1.4GHz} \sim L_{\rm X}^b$, $b=1.26$) which seems to
be steeper than the observed relation ($b \leq 1$). This would imply that if radio halos
are of hadronic origin, the parent cosmic ray proton population needs
to have a spatially wider energy density distribution than the thermal
gas.
\item The expected radio polarization is of the order of $10^{-6}$. This
is mainly a property of the magnetic field configuration, and should
also hold for other radio halo models.
\item The expected gamma ray fluxes are below present observational
limits, but possibly in the range of future instruments.
\item The set of 10 simulated clusters reproduces the observed radio
luminosity-temperature relation of clusters of galaxies surprisingly well.
\end{enumerate}

\begin{acknowledgements}
We thank Bruno Deiss, Wilhelm Reich, Harald Lesch, and Richard
Wielebinski for the access to the radio map of Coma. We also want to
thank Haida Liang for the access to her collected data of the
radio/x-ray luminosity relation, and Henk Spruit and Sabine Schindler
for carefully reading the manuscript.  Finally, we like to acknowledge
many helpful comments by Gabriele Giovannini, the referee.
\end{acknowledgements}

\def\aj{AJ}

\bibliography{tae,bib.kd}
\bibliographystyle{aabib99}

\end{document}